\documentclass[runningheads]{llncs}

\usepackage{graphicx}
\usepackage{amsmath}
\begin{document}

\title{CAR-Net: Unsupervised Co-Attention Guided Registration Network for Joint Registration and Structure Learning}


%
\author{Xiang Chen\inst{1} \and
Yan Xia\inst{1,2} \and
Nishant Ravikumar\inst{1,2} \and
Alejandro F Frangi\inst{1,2,3,4}}
\authorrunning{X. Chen et al.}
%
\institute{Center for Computational Imaging and Simulation Technologies in Biomedicine, School of Computing, University of Leeds, Leeds,UK \and
Biomedical Imaging Department, Leeds Institute for Cardiovascular and Metabolic Medicine, School of Medicine University of Leeds, Leeds, UK\\
\and Department of Cardiovascular Sciences, KU Leuven, Leuven, Belgium\\
\and
Department of Electrical Engineering, KU Leuven, Leuven, Belgium}

%
\maketitle              
\sloppy
\begin{abstract}
Image registration is a fundamental building block for various applications in medical image analysis. To better explore the correlation between the fixed and moving images and improve registration performance, we propose a novel deep learning network, Co-Attention guided Registration Network (CAR-Net). CAR-Net employs a co-attention block to learn a new representation of the inputs, which drives the registration of the fixed and moving images. Experiments on UK Biobank cardiac cine-magnetic resonance image data demonstrate that CAR-Net obtains higher registration accuracy and smoother deformation fields than state-of-the-art unsupervised registration methods, while achieving comparable or better registration performance than corresponding weakly-supervised variants. In addition, our approach can provide critical structural information of the input fixed and moving images simultaneously in a completely unsupervised manner. 

\end{abstract}
\section{Introduction}
Image registration is an active research area in computer vision and medical image analysis. It is central to facilitating numerous downstream tasks (e.g. anatomical motion tracking/deformation analysis, disease diagnosis and monitoring progression, among others). Traditional methods view this task as an iterative process to optimise the parameters of the deformation model under the guidance of dissimilarity/similarity metrics. Therefore, despite achieving high accuracy and possessing favourable geometric attributes, traditional registration methods are generally time-consuming. To speed up image registration, recently, deep learning (DL)-based methods have been widely applied in this domain, where the inference is just one forward pass through the network.

DL-based registration methods could be roughly divided into three categories, supervised, unsupervised and weakly-supervised methods. Supervised methods~\cite{rohe2017svf,eppenhof2018pulmonary,eppenhof2019progressively,cao2017deformable} can achieve comparable registration performance to traditional methods, but require only a fraction of time (relative to the latter) to register images during inference. However, the ground-truth deformation fields are challenging to obtain, which makes it an obstacle for large-scale training of DL-based methods. This problem can be overcome through unsupervised methods ~\cite{balakrishnan2019voxelmorph,dalca2019unsupervised,krebs2019learning}, which generally comprise an end-to-end convolutional neural network (CNN) to predict the deformation fields and a spatial transformer network (STN)~\cite{jaderberg2015spatial} to warp the moving image towards the fixed image. Dissimilarity/similarity between the warped moving image and the fixed image, and regularisation of the deformation fields constitute the loss function used to train networks. Although the ground-truth deformation fields are difficult to obtain, other types of auxiliary information (e.g. segmentation masks or landmarks) could also be used to guide registration. Weakly-supervised registration methods are based on introducing such `weak' labels, and have been shown to achieve better registration performance than unsupervised methods~\cite{balakrishnan2019voxelmorph,dalca2019unsupervised,hu2018label,mansilla2020learning}.

However, most previous DL-based registration networks focus on building an end-to-end network from the pair of moving and fixed images to estimate deformation fields, seldom considering if the input pairs are sufficiently descriptive for the task. To tackle this issue, Lee et al.~\cite{lee2019image} proposed an Image-and-Spatial Transformer Network (ISTN), containing a dedicated image transformer network (ITN) to generate new representations of the inputs, and an STN to register the new representations of the inputs, achieving better performance than using STN to register the original inputs directly. However, ISTN uses a shared CNN to extract new representations of the input images, which requires annotations (e.g. segmentations, landmarks or centerlines) and thus does not exploit the intrinsic correlation between them in an unsupervised manner. Several recent weakly-supervised approaches have proposed to first register the segmentation masks~\cite{chen2020semantic} or semantic features~\cite{ha2020semantically} at a global level instead of directly aligning the original input images, which also highlights the importance of introducing additional structural information. However, sourcing such additional annotations can be expensive or infeasible for specific applications, limiting the generality of such weakly-supervised methods.

Image registration always involves two or more images as inputs (to be co-registered), where the contents in these images are highly-correlated but may comprise differences in spatial/shape characteristics. We argue that highlighting the correlated regions in the moving and fixed images can help learn coarse structural information from the input images automatically. To minimise the impact of irrelevant pixels and improve registration performance, a co-attention guided registration network is proposed in this paper. The co-attention block is used to generate new representations of the input images, where, correlated pixels are highlighted and irrelevant context is dismissed. With the learned new representations, the performance of DL-based registration networks can be further improved. Additionally, following training on a large-scale dataset, the representations learned in an unsupervised manner, can be visualised as a coarse prediction of correlated structural information.

In summary, there are three main contributions of our proposed method: (1) a 3D co-attention block is designed to exploit the correlation of input image pairs to be registered, which can be easily incorporated into other networks. (2) Using the co-attention block, we build a new DL-based registration network, CAR-Net, which achieves state-of-the-art registration performance on cine-cardiac magnetic resonance (CMR) images from UK Biobank Imaging Study (UKBB). (3) Our co-attention block is shown to provide coarse structural information in an unsupervised manner, which is of potential value for developing explainable AI solutions to medical image registration.

\begin{figure}[h]
\begin{center}
\includegraphics[width=0.8\textwidth]{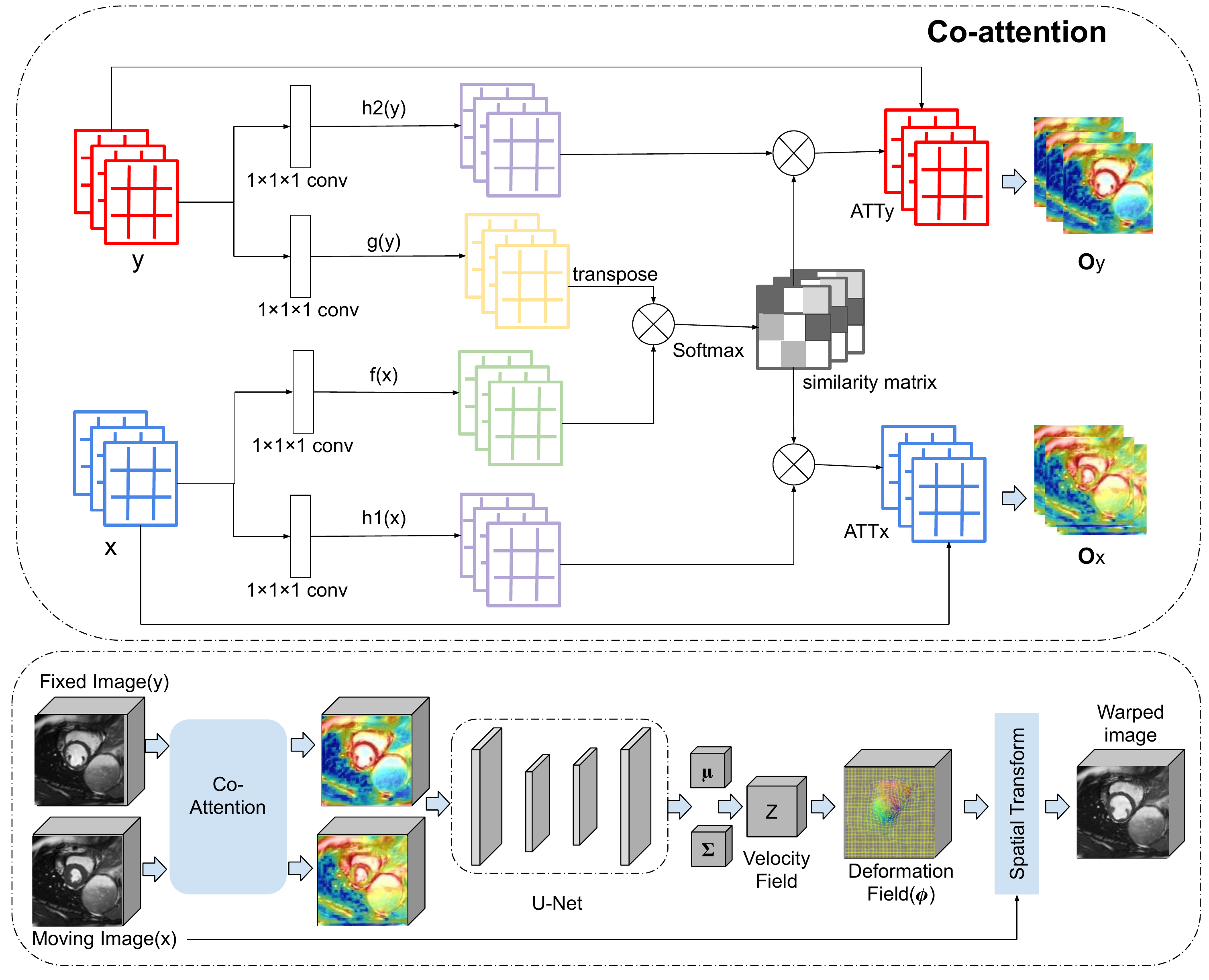}
\caption{Schema of our proposed CAR-Net. The whole registration network is displayed in the bottom row, with the details of the co-attention block in the top row. The co-attention block takes feature maps from the moving and fixed image as input (denoted as $x$ and $y$). It predicts corresponding attention maps (denotes as $ATT_x$ and $ATT_y$) and outputs ($\textbf{O}_{x}$ and  $\textbf{O}_{y}$), with the same size as inputs. The CMR images were reproduced with the permission of UK Biobank©.}
\label{fig:network}
\end{center}
\end{figure}

\section{Methods}
Given the moving image $\textbf{I}_M$ and fixed image $\textbf{I}_F$, image registration aims to find a coordinate mapping function $ \phi(\circ)$ from $\textbf{I}_M$ to $\textbf{I}_F$, formulated as,
\begin{equation}
\label{eqn:registration}
\begin{split}
\textbf{I}_F = \phi(\textbf{I}_M).
\end{split}
\end{equation}

Unsupervised DL-based image registration methods~\cite{balakrishnan2019voxelmorph,dalca2019unsupervised} initially concatenate $\textbf{I}_M$ and $\textbf{I}_F$, following which, an encoder-decoder is applied to predict the deformation fields. Finally, a STN block is used to deform the $\textbf{I}_M$ to $\textbf{I}_F$ based on the learned deformation fields. Networks are trained by optimising a loss function that computes a dissimilarity/similarity metric between the warped moving image and $\textbf{I}_F$, and regularises the estimated deformation fields. Our approach, CAR-Net, utilises each of these fundamental building blocks of unsupervised registration networks, and incorporates a co-attention block that learns new image representations from the original inputs automatically.

\textbf{Co-attention Block.} To learn the transformation from the moving image to the fixed image, the key is to establish correspondence between these two images. Therefore, pixels across different spatial positions in these two images do not share the same importance for registration. In fact, individual pixels/regions in the images may contain confounding information that can adversely impact registration accuracy. To limit the impact of such pixels/regions on overall registration accuracy, it is intuitive to focus the attention of the registration algorithm on the key points/structures in the moving and fixed images. To this end, we propose a co-attention block to exploit the correlations between the input moving and fixed images, for registration (as shown in Fig.~\ref{fig:network}).

The co-attention block is inspired by~\cite{lu2019see,wei2018fast,lu2016hierarchical}, where it is employed for question-answering and video segmentation tasks. It is however, well-suited to the task of image registration as well (as there are two highly correlated inputs, the moving and fixed image). Our co-attention block takes both features from moving and fixed images as input, and predicts the corresponding co-attention feature maps, where the highly-correlated and motion-related pixels in both feature maps are highlighted. The moving and fixed image feature maps $\textbf{F}_{mov},\textbf{F}_{fix} \in \mathcal{R}^{ W \times H \times D \times C}$ (C, W, H and D are channels, width, height and depth of feature maps) are first transformed into two different feature spaces by two corresponding $1\times1\times1$ convolution layers (denoted as $f(\circ)$ and $g(\circ)$) and flattened (from $\mathcal{R}^{W \times H \times D \times C}$ to $\mathcal{R}^{N\times C}, N=W \times H \times D$) to calculate similarity matrix $\textbf{S} \in \mathcal{R}^{N\times N}$. Based on similarity matrix $\textbf{S}$ and feature maps learned with two additional $1\times1\times1$ convolution layers $h_1(\circ)$ and $h_2(\circ)$ from the input, the fixed attention maps $\textbf{ATT}_{fix}$ and moving attention maps $\textbf{ATT}_{mov}$ are computed. The process of co-attention can be formulated as,
\begin{equation}
\label{eqn:co-attention}
\begin{aligned}
& \textbf{S} = f(\textbf{F}_{mov}) \times g(\textbf{F}_{fix})^T,\\
& \textbf{ATT}_{mov} = Softmax(\textbf{S}) \times h_2(\textbf{F}_{fix}),\\
& \textbf{ATT}_{fix} = Softmax(\textbf{S}^T) \times h_1(\textbf{F}_{mov}),\\
& \textbf{O}_{mov} = h_1(\textbf{F}_{mov}) + \alpha_1 \sigma(\textbf{ATT}_{mov}) \cdot h_1(\textbf{F}_{mov}), \\
& \textbf{O}_{fix} = h_2(\textbf{F}_{fix})+ \alpha_2 \sigma(\textbf{ATT}_{fix}) \cdot h_1(\textbf{F}_{fix}),
\end{aligned}
\end{equation}
where, $\textbf{O}_{mov}$ and $\textbf{O}_{fix}$ are the output feature maps (learned new representations) of $\textbf{F}_{mov}$ and $\textbf{F}_{fix}$ after co-attention respectively. $Softmax(\circ)$ is the Softmax function, applied to the last channel of the similarity matrix $\textbf{S}$. $\sigma(\circ)$ denotes the Sigmoid function, which is a $1\times 1 \times 1$ convolution layer followed by a Sigmoid activation. The $\alpha_1$ and $\alpha_2$ are weights for the attention maps, learned automatically by back-propagation. 


\textbf{Network Architecture.} CAR-Net can be seen as a combination of a co-attention block with a generic unsupervised registration network. The co-attention block learns new representations ($\textbf{O}_{mov}$ and $\textbf{O}_{fix}$) of the input moving and fixed images, which serve as inputs to the registration block that predicts deformation fields and yields warped moving images. The co-attention block can be directly applied to the input moving and fixed images, however, we apply two down-sampling layers before the co-attention block and an up-sampling block after it, due to computational constraints (GPU memory).
The parameters of the co-attention block and registration block are optimised using the same loss function, allowing the co-attention block to learn the most relevant/correlated structural features from the input images in an unsupervised manner. 

To ensure the generated deformation fields are smooth, we parameterise deformation fields $\phi$ by stationary velocity fields (SVF) $z$ following previous research~\cite{krebs2019learning,dalca2019unsupervised}.
The registration part is a probabilistic model, comprising two main sub-blocks, a U-Net~\cite{ronneberger2015u} like structure and a spatial transform block. The former is an encoder-decoder network (three down-sampling layers and three up-sampling layers) to estimate deformation fields $\phi_z$, while the latter is used to deform the moving image to the fixed image based on the learned deformation fields from U-Net. The U-Net takes the concatenation of fixed and moving features [$\textbf{F}_{mov}$,$\textbf{F}_{fix}$,$\textbf{O}_{mov},\textbf{O}_{fix}]$ as input and predicts feature maps at $64 \times 64 \times 16$, with which mean $\mu$ and variance $\Sigma$ matrix are computed. Similar to~\cite{dalca2019unsupervised}, the velocity field $z$ is sampled from a multivariate Gaussian distribution using the estimated $\mu$ and $\Sigma$, followed by an integration layer and up-sampling layer to predict a diffeomorphic deformation field $\phi_z$. The predicted deformation field and moving image are fed into the spatial transform block (more details can be found in~\cite{jaderberg2015spatial}) to predict the warped moving image. 

\textbf{Loss Functions.} The loss function comprises two terms, the similarity loss and the regularisation of the deformation fields. In this paper, we choose normalised cross-correlation (NCC) as the similarity loss $L_s$, to evaluate the similarity between the warped moving image and the fixed image. 
Given warped moving image $x$ and fixed image $y$, $L_s$ is computed as,
\begin{equation}
\label{eqn:similarity}
{L}_{s} = 1- \frac{\sum_{i} (x_i-x_m)(y_i-y_m)}{\sqrt{\sum_{i}(x_i-x_m)^2} \sqrt{\sum_{i}(y_i-y_m)^2}}, 
\end{equation}
where, $x_i$, $y_i$ are the intensity of the i-th pixel in the warped moving image and fixed image respectively, and $x_m$, $y_m$ are the mean intensities in the corresponding images.

The deformation fields are regularised to improve the smoothness of predicted deformation fields. Following VM~\cite{dalca2019unsupervised}, we calculate the KL divergence between the approximate posterior $q_{\psi}(z|\textbf{I}_F;\textbf{I}_M)$ and the prior $p(z)$ ($p(z) = \mathcal N(z;0,\Sigma_{z})$) of velocity field $z$, formulated as,
\begin{equation}
\label{eqn:regularisation}
{L}_{r} = KL(q_{\psi}(z|\textbf{I}_F;\textbf{I}_M)||p(z|\textbf{I}_F;\textbf{I}_M)),
\end{equation}
where $q_{\psi}(z|\textbf{I}_F;\textbf{I}_M)$ is a multivariate normal,
\begin{equation}
\label{eqn:posterior}
q_{\psi}(z|\textbf{I}_F;\textbf{I}_M) = \mathcal N(z;\mu_{\textbf{z}|\textbf{I}_F,\textbf{I}_M},\Sigma_{\textbf{z}|\textbf{I}_F,\textbf{I}_M}),
\end{equation}
where $\mu_{\textbf{z}|\textbf{I}_F,\textbf{I}_M}$ and $\Sigma_{\textbf{z}|\textbf{I}_F,\textbf{I}_M}$ are the mean and variance of the distribution, learned by convolution layers.
Therefore, the total loss function is formulated as,
\begin{equation}
\label{eqn:total}
\begin{split}
{L}_{total} = \lambda_0 \times {L}_{s} + \lambda_1 \times L_r, 
\end{split}
\end{equation}
where, $\lambda_0$ and $\lambda_1$ are hyper-parameters that weight the relative influence of each loss term, which is to be defined in experiments. 

\section{Experiments and Results}
\textbf{Data and Implementation.} To demonstrate the performance of CAR-Net, we perform intra-subject registration experiments on cine CMR images (spatial resolution at $\sim 1.8 \times 1.8 \times 10 mm^3$) from UKBB. 2,000 subjects are randomly chosen from the UKBB dataset, each containing volumes at end-diastole (ED) and end-systole (ES), totally 4,000 volumes. These 2,000 subjects are then randomly split into a training set (1,600 subjects) and a testing set (400 subjects). We focus on intra-subject registration between ED and ES (i.e. ED-to-ES and ES-to-ED). A total of 3,200 image pairs were used for training, and 800 pairs for testing. Following previous cardiac registration methods~\cite{krebs2019learning}, we first resample all original CMR images into a spatial resolution at $1.50 \times 1.50 \times 3.15 mm^3$, then crop them into the same size of $128 \times 128 \times 32$ (using zero-padding if the slices are less than 32). To quantitatively evaluate the registration performance, we obtain segmentation masks for all volumes using the segmentation method proposed in~\cite{bai2018automated}. CAR-Net was implemented in Python using Keras, and trained using a single Tesla M60 GPU. The Adam optimiser, with a learning rate of 1e-4 was used to train our network. We set the batch size to 2 due to limited GPU memory. Hyper-parameters $\lambda_0$ and $\lambda_1$ were determined empirically, and set to 20 and 0.1, respectively. The source code will be publicly available on Github following review and acceptance of the paper.

\begin{figure}[h!]
\setlength{\belowcaptionskip}{-0.5 cm}
\begin{center}
\includegraphics[width=1.0\textwidth]{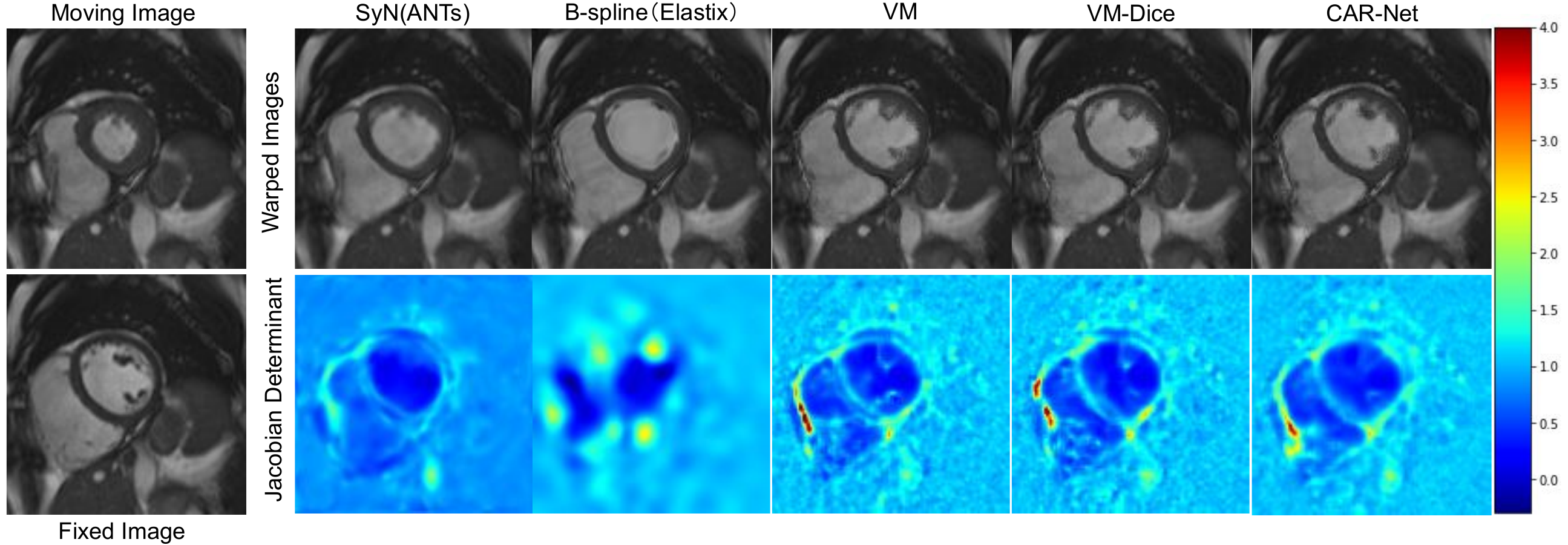}
\caption{Qualitative results for our CAR-Net and baseline networks viz. B-spline, SyN, VM and VM-Dice. The input moving and fixed images are on the left, while the corresponding warped moving images and Jacobian determinant of deformation fields (color bar is from -0.3 to 4) predicted by different approaches are on the right. The CMR images were reproduced with the permission of UK Biobank©.}
\label{fig:qualitative_results}
\end{center}
\end{figure}

\textbf{Comparison and Analysis.} We compared the performance of CAR-Net with traditional and state-of-the-art DL-based registration methods, using data from UKBB. For traditional methods, we choose the B-spline registration method (max iteration step is 2000, sampling 6000 random points per iteration) available in SimpleElastix~\cite{marstal2016simpleelastix} and Symmetric Normalisation (SyN)~\cite{avants2008symmetric} in ANTs (resolution level is 3 with 100 iterations in each sampling level) for comparison. For DL-based methods, we compare our CAR-Net with VM~\cite{dalca2019unsupervised} (using NCC as similarity loss) and a weakly-supervised version of VM (named VM-Dice), which uses Dice score as an additional loss to guide the network training. Note that, the main difference between CAR-Net and VM is the incorporation of co-attention block, while the other hyper parameter settings of the networks (e.g. convolution kernels, channels in the registration block and learning rate) are the same. 

\begin{table*}[h] 
 \caption{\label{tab:comparison} Quantitative comparison between CAR-Net and state-of-the-art methods using the Dice score of LVBP, LVM, RV, average Dice (denoted as Avg. Dice), HD and number of foldings (denoted as $|J_\phi|\le 0$).} 
 \centering
 \resizebox{12cm}{!}{
 \begin{tabular}{lcccccc} 
 \hline
  Methods  & LVBP Dice (\%) & LVM Dice (\%) & RV Dice (\%) & Avg. Dice (\%) & HD (mm) & $|J_\phi|\le 0$ \\ 
 \hline
  before Reg & $57.68 \pm 6.21$ & $30.88 \pm 8.68$ & $55.13 \pm 7.51$ & $47.90 \pm 6.33$ & $12.91 \pm 2.48$ & $-$\\ 
  B-spline & $74.44 \pm 11.50$ & $68.06 \pm 7.20$ & $61.76 \pm 12.05$ & $68.09 \pm 8.76$ & $13.72 \pm 3.57$ & $1671.97 \pm 1829.90$\\ 
  SyN & $70.92 \pm 9.36$ & $57.88 \pm 10.59$ & $60.30 \pm 8.35$ & $63.03 \pm 8.29$ & $12.98 \pm 2.68$ & $0.00 \pm 0.00$\\ 
  VM & $81.73 \pm 8.71$ & $72.04 \pm 4.65$ & $65.73 \pm 9.62$ & $73.16 \pm 6.26$ & $12.96 \pm 3.14$ & $0.29\pm 2.32$\\ 
  VM-Dice & $82.28 \pm 8.75$ & $72.53 \pm 4.59$ & $66.30 \pm 9.67$ & $73.70 \pm 6.28$ & $13.00 \pm 3.24$& $0.22 \pm 1.75$ \\
 \hline
 CAR-Net & $\textbf{82.59$\pm$ 8.35}$  & $\textbf{72.87$\pm$ 4.60} $ &   $\textbf{66.58$\pm$ 9.57} $ & $\textbf{74.01 $\pm$ 6.18}$ & $\textbf{12.73 $\pm$ 3.21}$ & $\textbf{0.00$\pm$ 0.00}$\\
 \hline
 \end{tabular}
 }
\end{table*}

\textbf{Qualitative and Quantitative Results.} The qualitative comparison between CAR-Net and state-of-the-art methods is shown in Fig.~\ref{fig:qualitative_results}. We find that, the prediction by SyN and B-spline only captures coarse deformation from the moving image to the fixed image. DL-based methods achieve better registration performance than traditional methods, while warped images predicted by CAR-Net are more consistent with the fixed image, with smoother deformations. 

Dice score for left ventricle blood pool (LVBP), left ventricle myocardium (LVM), right ventricle (RV) and average Dice), Hausdorff Distance (HD), and the number of foldings in the estimated deformation field (where Jacobian determinant is non-positive), are used to quantitatively evaluate the performance of our proposed method (as shown in Table\ref{tab:comparison}). Higher Dice score and lower HD stand for higher registration accuracy, while fewer foldings points to smoother deformation fields. 

Quantitative results are consistent with the visual assessment of the image registration using each approach. 
DL-based methods were found to achieve much higher Dice scores than traditional registration methods (of 5-10\% improvement). Our proposed CAR-Net consistently outperforms VM and other state-of-the-art methods in all metrics, achieving a higher Dice score ($\sim 1\%$ improvement than VM), lower HD and fewer foldings in the deformation fields. Note that, CAR-Net achieves completely smooth deformation, without foldings in the deformation fields. 
According to statistical analysis, the incorporation of a co-attention block (CAR-Net) provides statistically significant improvements over VM (i.e. without co-attention) in terms of the Dice scores for LVBP, LVM and the average (P-value$<$0.05). Although CAR-Net offers no significant improvement over the weakly-supervised version of VM (VM-Dice), CAR-Net is totally unsupervised and hence is not dependent on the availability of segmentation masks for structures/regions of interest. 

\textbf{Registration on Separate Regions.} To capture the motion of specific regions, it is sometimes necessary to register them independently (e.g. LVBP, LVM, RV). Using the same trained models, we also compare the performance of CAR-Net and VM on the registration of separate LVBP, LVM and RV regions (generated based on the same testing set), as shown in Table\ref{tab:sub_comparison}. Our CAR-Net achieves statistically significant improvements over both VM and VM-Dice in all metrics (Dice and HD) except the Dice score on LVBP. This further demonstrates the incorporation of co-attention block would bring more robust and accurate registration. 
A further visualisation and analysis of the learned co-attention maps is included in the supplementary material.

\begin{table*}[h] 
 \caption{\label{tab:sub_comparison} Quantitative comparison between CAR-Net and VM on registration of sub regions (LVBP, LVM and RV) of cardiac images. Statistical significant improvements compared with other methods are highlighted with bold (P-value$<$0.05).} 
 \centering
 \resizebox{12cm}{!}{
 \begin{tabular}{lccccccccc} 
 \hline
  Methods  & LVBP Dice (\%) & LVBP HD (mm)& LVBP $|J_\phi|\le 0$ & LVM Dice (\%)& LVM HD (mm)& LVM $|J_\phi|\le 0$ & RV Dice (\%) & RV HD (mm) & RV $|J_\phi|\le 0$ \\ 
 \hline
  before Reg & $57.68 \pm 6.21$ & $9.91 \pm 1.98$ & $-$ & $30.88 \pm 8.68$ & $9.00 \pm 1.25$ & $-$ & $55.13 \pm 7.51$ & $12.87 \pm 2.37$ & $-$\\ 
  VM & $88.98 \pm 4.93$ & $5.11 \pm 2.52$ & $0.00 \pm 0.11$ & $61.97 \pm 9.91$ & $9.36 \pm 1.74$ & $0.00\pm 0.04$& $81.42 \pm 6.80$ & $9.92 \pm 3.36$ & $0.15\pm 1.91$\\
  VM-Dice & $90.16 \pm 4.71$ & $4.73 \pm 2.56$ & $0.00 \pm 0.05$ & $64.23 \pm 9.08$ & $9.66 \pm 1.84$& $0.00 \pm 0.00$ & $83.75 \pm 5.85$ & $9.25 \pm 3.17$& $0.04 \pm 0.68$\\
 \hline
 CAR-Net & $90.65\pm 3.91$  & $\textbf{4.17$\pm$ 2.29} $ &   $0.00\pm 0.00 $ & $\textbf{71.55 $\pm$ 6.68}$ & $\textbf{8.13 $\pm$ 2.03}$ & $0.00\pm 0.00$& $\textbf{86.31 $\pm$ 5.02}$ & $\textbf{8.17 $\pm$ 3.13}$ & $\textbf{0.00$\pm$ 0.00}$\\
 \hline
 \end{tabular}
 }
\end{table*}

\section{Conclusion}
This paper proposes a novel DL-based image registration network, CAR-Net, for 3D CMR image registration. The key idea is to incorporate 3D co-attention into a generic unsupervised registration framework, to learn highly-correlated structural information between the moving and fixed images automatically. Experiments on UKBB cardiac images demonstrate that CAR-Net achieves higher registration accuracy, completely smooth deformation fields and is more robust than state-of-the-art unsupervised methods and a weakly-supervised version of VM. 
CAR-Net employs co-attention to learn new representations of input images automatically, which could be seen as a coarse prediction of essential structural information and therefore, lends itself to interpreting network decisions. 


\bibliographystyle{splncs04} 
\bibliography{coattention}

\end{document}